\documentclass[a4paper]{article}

\usepackage[english]{babel}
\usepackage{amsmath}
\usepackage[symbol]{footmisc}
\usepackage{amssymb}
\usepackage{fancyhdr}
\usepackage[margin=1in]{geometry} 

\usepackage{graphicx}
\usepackage[export]{adjustbox}
\usepackage{enumitem}
\usepackage{textcase}
\usepackage{multicol}
\usepackage{xcolor}
\usepackage[tableposition=top,font=small,labelfont=bf]{caption}
\usepackage{booktabs}
\usepackage{siunitx}
\usepackage{subcaption}
\usepackage{floatrow}
\usepackage{comment}
\usepackage{multirow}
\usepackage{hhline}
\usepackage{gensymb}
\usepackage{rotating}
\usepackage{makecell}
\usepackage[breaklinks]{hyperref}

\usepackage{algorithm, algpseudocode} 
\usepackage{mathrsfs} 
\usepackage{mathptmx}
\usepackage{sidecap}
\usepackage{comment}
\usepackage{ragged2e}
\usepackage{wasysym}
\usepackage{soul}

\usepackage{amsmath}
\usepackage{amsthm}
\usepackage{amssymb}
\usepackage{booktabs}
\usepackage{authblk}

\usepackage[utf8]{inputenc}
\usepackage{hyperref}
\usepackage{url}
\hypersetup{
	unicode,
	pdfauthor={Author One, Author Two, Author Three},
	pdftitle={A simple article template},
	pdfsubject={A simple article template},
	pdfkeywords={article, template, simple},
	pdfproducer={LaTeX},
	pdfcreator={pdflatex}
}

\setitemize{font=\scriptsize}

\numberwithin{equation}{section}

\definecolor{lightgray}{gray}{0.9}

\usepackage{pdfpages}
\usepackage{multicol}

\usepackage[sorting=none, style=ieee]{biblatex} 
\bibliography{refs}
\usepackage{sectsty}
\usepackage{soul}
\usepackage{setspace}
\sectionfont{\large}
\usepackage{graphicx, color}
\usepackage{orcidlink}
\graphicspath{{fig/}}

\title{\vspace{-2cm}Decoupling Pulse Tube Vibrations from a Dry Dilution Refrigerator at milli-Kelvin Temperatures}

\usepackage[utf8]{inputenc}

\usepackage{authblk}

\usepackage{orcidlink}

\newcommand{\NUCLEUS}{}

\newcommand\extrafootertext[1]{%
    \bgroup
    \renewcommand\thefootnote{\fnsymbol{footnote}}%
    \renewcommand\thempfootnote{\fnsymbol{mpfootnote}}%
    \footnotetext[0]{#1}%
    \egroup
}

 \newcommand{\TUW}{1}
    \newcommand{\MPP}{2}
    \newcommand{\INFNRoma}{3}
    \newcommand{\INFNTorVergata}{4}
    \newcommand{\CNR}{5}
    \newcommand{\Sapienza}{6}
    \newcommand{\TUM}{7}
    \newcommand{\HEPHY}{8}
    \newcommand{\CEA}{9}
    \newcommand{\Ferrara}{10}
    \newcommand{\IrfuAPC}{11}
    \newcommand{\LNGS}{12}
    \newcommand{\TorVergata}{13}
    \newcommand{\INFNFerrara}{14}
    \newcommand{\Coimbra}{15}
    \newcommand{\Bicocca}{16}
    \newcommand{\MPIK}{17}

\date{}
\author[\TUM]{A.~Wex\footnote{Corresponding author. E-mail: \textcolor{blue}{alexander.wex@tum.de}}~\orcidlink{0009-0003-5371-2466}}
\author[\TUM]{J.~Rothe}
\author[\TUM, \CEA]{L.~Peters}
\author[\TUW]{\\\vspace{0.25em}H.~Abele~\orcidlink{0000-0002-6832-9051}}
\author[\MPP]{G.~Angloher}
\author[\HEPHY]{B.~Arnold}
\author[\TorVergata, \INFNTorVergata]{M.~Atzori~Corona}
\author[\MPP, \Coimbra]{A.~Bento~\orcidlink{0000-0002-3817-6015}}
\author[\CEA]{E.~Bossio~\orcidlink{0000-0001-9304-1829}}
\author[\HEPHY]{J.~Burkhart~\orcidlink{0000-0002-1989-7845}}
\author[\MPP, \thanks{Now at: \Bicocca}]{L.~Canonica~\orcidlink{0000-0001-8734-206X}}
\author[\INFNRoma]{F.~Cappella~\orcidlink{0000-0003-0900-6794}}
\author[\Sapienza, \INFNRoma]{M.~Cappelli~\orcidlink{0009-0002-6148-5964}}
\author[\INFNRoma]{N.~Casali~\orcidlink{0000-0003-3669-8247}}
\author[\TorVergata, \INFNTorVergata]{R.~Cerulli~\orcidlink{0000-0003-2051-3471}}
\author[\INFNRoma]{A.~Cruciani~\orcidlink{0000-0003-2247-8067}}
\author[\Sapienza, \INFNRoma]{G.~Del~Castello~\orcidlink{0000-0001-7182-358X}}
\author[\Sapienza, \INFNRoma]{M.~del~Gallo~Roccagiovine}
\author[\TUW]{A.~Doblhammer}
\author[\TUW]{S.~Dorer~\orcidlink{0009-0001-1670-5780}}
\author[\TUM]{A.~Erhart~\orcidlink{0000-0002-8721-177X}}
\author[\HEPHY]{M.~Friedl~\orcidlink{0000-0002-7420-2559}}
\author[\HEPHY]{S.~Fichtinger}
\author[\MPP]{A.~Garai}
\author[\HEPHY]{V.M.~Ghete~\orcidlink{0000-0002-9595-6560}}
\author[\TorVergata, \INFNTorVergata]{M.~Giammei~\orcidlink{0009-0006-9104-2055}}
\author[\CEA, \thanks{Now at: \MPIK}]{C.~Goupy~\orcidlink{0000-0003-4954-5311}}
\author[\MPP, \TUM]{D.~Hauff}
\author[\CEA]{F.~Jeanneau~\orcidlink{0000-0002-6360-6136}}
\author[\TUW]{E.~Jericha~\orcidlink{0000-0002-8663-0526}}
\author[\TUM]{M.~Kaznacheeva~\orcidlink{0000-0002-2712-1326}}
\author[\TUM]{A.~Kinast~\orcidlink{0000-0002-5894-2303}}
\author[\HEPHY]{H.~Kluck~\orcidlink{0000-0003-3061-3732}}
\author[\MPP]{A.~Langenk\"{a}mper}
\author[\IrfuAPC, \TUM]{T.~Lasserre~\orcidlink{0000-0002-4975-2321}}
\author[\CEA]{D.~Lhuillier~\orcidlink{0000-0003-2324-0149}}
\author[\MPP]{M.~Mancuso~\orcidlink{0000-0001-9805-475X}}
\author[\CEA, \TUW]{R.~Martin}
\author[\MPP]{B.~Mauri}
\author[\INFNFerrara]{A.~Mazzolari}
\author[\CEA]{E.~Mazzucato}
\author[\CEA]{H.~Neyrial}
\author[\CEA]{C.~Nones}
\author[\TUM]{L.~Oberauer}
\author[\TUM]{T.~Ortmann}
\author[\LNGS, \thanks{Now at: \Bicocca}]{L.~Pattavina~\orcidlink{0000-0003-4192-849X}}
\author[\MPP]{F.~Petricca~\orcidlink{0000-0002-6355-2545}}
\author[\TUM]{W.~Potzel}
\author[\MPP]{F.~Pr\"{o}bst}
\author[\MPP]{F.~Pucci}
\author[\HEPHY, \TUW]{F.~Reindl~\orcidlink{0000-0003-0151-2174}}
\author[\INFNFerrara]{M.~Romagnoni}
\author[\TUM]{N.~Schermer~\orcidlink{0009-0004-4213-5154}}
\author[\HEPHY, \TUW]{J.~Schieck~\orcidlink{0000-0002-1058-8093}}
\author[\TUM]{S.~Sch\"{o}nert~\orcidlink{0000-0001-5276-2881}}
\author[\HEPHY, \TUW]{C.~Schwertner}
\author[\CEA]{L.~Scola}
\author[\CEA]{G.~Soum-Sidikov~\orcidlink{0000-0003-1900-1794}}
\author[\MPP]{L.~Stodolsky}
\author[\TUM]{R.~Strauss~\orcidlink{0000-0002-5589-9952}}
\author[\Ferrara, \INFNFerrara]{M.~Tamisari}
\author[\HEPHY]{R.~Thalmeier~\orcidlink{0009-0003-4480-0990}}
\author[\INFNRoma]{C.~Tomei}
\author[\Sapienza, \INFNRoma]{M.~Vignati~\orcidlink{0000-0002-8945-1128}}
\author[\CEA]{M.~Vivier~\orcidlink{0000-0003-2199-0958}}
\author[\TUM]{V.~Wagner~\orcidlink{0000-0003-1845-4951}}

\affil[\TUW]{Atominstitut, Technische Universit\"at Wien, A-1020 Wien, Austria \NUCLEUS}
    
\affil[\MPP]{Max-Planck-Institut f\"ur Physik, D-80805 M\"unchen, Germany \NUCLEUS}
    
\affil[\INFNRoma]{Istituto Nazionale di Fisica Nucleare -- Sezione di Roma, Roma I-00185, Italy}
    
\affil[\INFNTorVergata]{Istituto Nazionale di Fisica Nucleare -- Sezione di Roma "Tor Vergata", Roma I-00133, Italy \NUCLEUS}
    
\affil[\CNR]{Consiglio Nazionale delle Ricerche, Istituto di Nanotecnologia, Roma I-00185, Italy \NUCLEUS}
    
\affil[\Sapienza]{Dipartimento di Fisica, Sapienza Universit\`{a} di Roma, Roma I-00185, Italy \NUCLEUS}
    
\affil[\TUM]{Physik-Department, Technische Universit\"at M\"unchen, D-85748 Garching, Germany \NUCLEUS}
    
\affil[\HEPHY]{Institut f\"ur Hochenergiephysik der \"Osterreichischen Akademie der Wissenschaften, A-1050 Wien, Austria \NUCLEUS}
    
\affil[\CEA]{IRFU, CEA, Universit\'{e} Paris-Saclay, F-91191 Gif-sur-Yvette, France \NUCLEUS}
    
\affil[\Ferrara]{Dipartimento di Fisica, Universit\`{a} di Ferrara, I-44122 Ferrara, Italy \NUCLEUS}
    
\affil[\IrfuAPC]{IRFU (DPhP \& APC), Universit\'{e} Paris-Saclay, F-91191 Gif-sur-Yvette, France \NUCLEUS}
    
\affil[\LNGS]{Istituto Nazionale di Fisica Nucleare -- Laboratori Nazionali del Gran Sasso, Assergi (L’Aquila) I-67100, Italy \NUCLEUS}
    
\affil[\TorVergata]{Dipartimento di Fisica, Universit\`{a} di Roma "Tor Vergata", Roma I-00133, Italy \NUCLEUS}
    
\affil[\INFNFerrara]{Istituto Nazionale di Fisica Nucleare -- Sezione di Ferrara, I-44122 Ferrara, Italy \NUCLEUS}
    
\affil[\Coimbra]{LIBPhys-UC, Departamento de Fisica, Universidade de Coimbra, P3004 516 Coimbra, Portugal\NUCLEUS}
    
\affil[\Bicocca]{Dipartimento di Fisica, Universit\`{a} di Milano Bicocca, 20126, Milan, Italy}

\affil[\MPIK]{Max-Planck-Institut für Kernphysik, D-69117 Heidelberg,
Germany}

\begin{document}

	\maketitle

\vspace{-4em}

	\begin{abstract}
    With the rising adoption of dry dilution refrigerators across scientific and industrial domains, there has been a pressing demand for highly efficient vibration decoupling systems capable of operation at cryogenic temperatures in order to achieve the low vibration levels required for operation of sensitive equipment like cryogenic detectors or quantum devices. As part of the NUCLEUS experiment, a cryogenic spring pendulum has been engineered to effectively isolate pulse tube vibrations by establishing an autonomous frame of reference for the experimental volume, while sustaining temperatures below 10\,mK. Attaining attenuation of up to two orders of magnitude within the region of interest of the NUCLEUS cryogenic detectors, we achieved displacement RMS values in the order of 1\,nm in the axial direction and 100\,pm radially, thereby reducing vibrations below typical environmental levels. Our successful detector operation across multiple cooldown cycles demonstrated negligible sensitivity to pulse tube induced vibrations, culminating in the achievement of an ultra-low $(6.22 \pm 0.07)$\,eV baseline resolution on a gram-scale CaWO$_4$ cryogenic calorimeter during continuous pulse tube operation over the course of several weeks.
\vspace{1em}
		
	\end{abstract}
    \twocolumn
    \tableofcontents


    \section{Introduction}
\label{sec:intro}
The growing popularity of quantum devices requiring ultra-low temperatures for optimal performance, is driving a steady increase in the cryostat market in recent years \cite{Cryo_market}. Among the various technologies available, dilution refrigerators achieving base temperatures below 10\,mK are essential. While wet dilution refrigerators, which require a continuous supply of liquid helium, have traditionally met this demand, the recent development of dry dilution refrigerators (DDRs) offers an increasingly favored alternative. These advancements are not only pivotal for the field of quantum technology but also for physics experiments searching for rare events like neutrino-less double-beta decay (e.g. CUORE \cite{adams2022cuore}, CUPID \cite{armengaud2021new}, AMoRE \cite{jo2018status}), dark matter search (e.g. CRESST \cite{abdelhameed2019first}, SuperCDMS \cite{aralis2020constraints}, EDELWEISS \cite{arnaud2020first}) and detection of coherent elastic neutrino-nucleus scattering, or CE$\nu$NS (e.g. NUCLEUS \cite{Angloher:2019flc}, RICOCHET \cite{augier2023ricochet}, MINER \cite{agnolet2017background}). \\
While dry dilution refrigerators offer the possibility of operating at temperatures below 10\,mK for an unlimited time without external helium supply, this comes at the price of increased vibrations due to their pulse tube cooled 40\,K and 4\,K stages. Built around the principle of periodically compressing and expanding a gas volume their characteristic vibrations have frequencies of one to a few Hz \cite{ventura2010art,kalra2016vibration}, that are problematic for sensitive measurement equipment \cite{olivieri2017vibrations}. Efforts are taken by cryostat manufacturers to reduce the impact of these vibrations, for example by using bellows to mechanically decouple the pulse tube from the cryostat \cite{BF_LD,olivieri2017vibrations}. As pulse tube operation still introduces strong vibrations at mixing chamber level, many research experiments develop dedicated vibration decoupling systems (VDS) specially designed to their requirements \cite{de2019vibration,maisonobe2018experimental,pirro2006further}. \\
This paper details the development of a cryogenic VDS optimized for the operation of sensitive particle detectors in DDRs. The VDS was developed in the scope of the NUCLEUS experiment which aims to detect CE$\nu$NS at the Chooz nuclear power plant in France, by measuring the recoil energy in the scattering process of reactor anti-neutrinos on the nuclei of gram-scale CaWO$_4$ target crystals. Detection of the resulting recoil energies in the order of tens to hundreds of eV happens via transition edge sensors kept at operating temperatures around 10\,mK \cite{Angloher:2019flc}. The detectors exhibit a strong sensitivity to both environmental and pulse tube related vibrations that prevents operation directly on the mixing chamber stage. This motivated the development of a VDS that creates a detector environment cooled to mixing chamber temperature, mounted entirely on an independent frame of reference, which is described in the following.\\
The paper is outlined as follows. Section \ref{sec:spring_based_pendulum} provides an overview of the NUCLEUS VDS's components and gives a basic theoretical description of its features. Special emphasis is put on overcoming the thermal design challenge of creating a non-rigid direct connection between the VDS mounting point at room temperature and temperatures below 10\,mK. Section \ref{sec:experimental_results} quantitatively assesses the system's performance using vibration measurements performed on the setup at room temperature as well as results from cooldowns of the setup with detector operation in the scope of NUCLEUS R\&D runs. The paper is concluded in section \ref{sec:conclusion}.

 
 \section{Concept of a Spring-Based Cryo-Pendulum} 
    \label{sec:spring_based_pendulum}
    \subsection{Cryostat and Measurement Equipment}
    \label{sec:equipment}
    The NUCLEUS experiment uses a standard issue commercial DDR of the type LD400 produced by BlueFors. Achieved base temperature on the cryostat's mixing chamber (MC) lies below 8\,mK at a nominal cooling power of 17.5\,$\mu$W at 20\,mK. A Cryomech PT-415 pulse tube, operating at a 1.4\,Hz frequency, is used for cooling down to 4\,K. It is thermally coupled to the 40\,K and 4\,K stages using thick copper braids in order to limit transmission of vibrations to the cryostat \cite{BF_LD, kalra2016vibration}.  \\    
    For a quantitative comparison of vibrational noise conditions, accelerometer measurements were performed using a PCB Piezotronics 393B05 with 10.26 V/g sensitivity. All vibration measurements presented in the following were conducted at room temperature with the cryostat's vacuum chamber closed, vacuum pumps operating and pressure reduced below 1\,mbar to simulate realistic operating conditions. The accelerometer data is converted into the frequency domain using Welch's method \cite{welch1967use}. Conversion into displacement units is done using \cite{olivieri2017vibrations}:
    \begin{equation}
    \label{eq:displacement}
        PSD_d(f) = \frac{g}{(2\pi\cdot f)^2} \cdot PSD_a (f) \, .
    \end{equation}
    Here $PSD_d$ denotes the power spectral density in units of displacement ($m/\sqrt{Hz}$) and $PSD_a$ in units of acceleration ($g/\sqrt{Hz}$). $f$ is the corresponding frequency and $g$ the gravitational acceleration. By integrating \ref{eq:displacement} over chosen frequency ranges, quantitative comparisons between different setups can be performed. The main focus is the NUCLEUS cryogenic detectors' region of interest (ROI) between 15\,Hz and 1000\,Hz, which is defined as the frequency range contributing 90\,\% weight to the detector resolution and is derived from the detectors' signal and noise spectral densities \cite{rothe2021low}. To provide a full picture, we also integrate the region below ROI and the full frequency range up to the sensor's sensitivity limit around 2\,kHz.\\
    To quantify pulse tube influence on the vibration levels, accelerometer measurements were performed on the mixing chamber stage. Root mean square (RMS) of accelerometer data computed over different frequency regions are listed in table \ref{tab:axial_benchmark}\footnote{For reasons of comparability the displacement spectra are only integrated inside the region of interest, as the diverging behavior towards $f = 0$ causes the results to be strongly dependent on the chosen cut-off frequency.}. During pulse tube operation the integrated displacement amounts to values in the hundreds of nanometers in axial and tens of nanometers in radial direction. This corresponds to a factor 2 increase in axial and a factor 5 increase in radial direction by pulse tube operation. When testing the operation of NUCLEUS-type TES detector directly on the mixing chamber it was found that they are mostly too disturbed by pulse tube vibrations for any stable operation. In contrast, briefly turning off the pulse tube allowed short-time detector operation at nominal performance. This stresses the necessity of an efficient vibration decoupling solution achieving vibrations level comparable or below what is reached on the mixing chamber with pulse tube turned off. \\

   \begin{table*}[htbp]
    \centering
    \begin{tabular}{|c|c|c|c|c|}
    \hline
       Frequency range &  \multicolumn{2}{c|}{DDR PT off}& \multicolumn{2}{c|}{DDR PT on} \\
       \hline
        & Axial & Radial & Axial & Radial\\
        \hline
        [Hz]&  \multicolumn{2}{c|}{Displacement [nm]} & \multicolumn{2}{c|}{Displacement [nm]} \\
       \hline
       
       \hline
        15-1000 & 87.5 $\pm$ 7.5 & 7.58 $\pm$ 0.65&174 $\pm$ 15 &38.4 $\pm$ 3.3\\ 
        \hline
        [Hz] &  \multicolumn{2}{c|}{Acceleration [mg]} & \multicolumn{2}{c|}{Acceleration [mg]} \\
       \hline
        0-15 & 0.057 $\pm$ 0.005 & 0.45 $\pm$ 0.04 & 0.047 $\pm$ 0.004 & 1.11 $\pm$ 0.09 \\ 
        15-1000 & 1.07 $\pm$ 0.09 & 0.22 $\pm$ 0.02 & 2.23 $\pm$ 0.19 & 1.09 $\pm$ 0.09\\ 
        0-2000 & 1.13 $\pm$ 0.10 & 0.67 $\pm$ 0.06& 2.32 $\pm$ 0.20& 2.20 $\pm$ 0.19 \\ 
        \hline
    \end{tabular}
    \caption{Benchmark accelerometer measurements performed on the mixing chamber plate. The uncertainties are calculated from the averaging of individual measurement windows performed in Welch's method.}
    \label{tab:axial_benchmark}
\end{table*}

   \subsection{Overview of the Vibration Decoupling System}
   \label{sec:system_description}
\begin{figure}[htbp]
    \centering
      \includegraphics[width=1\linewidth]{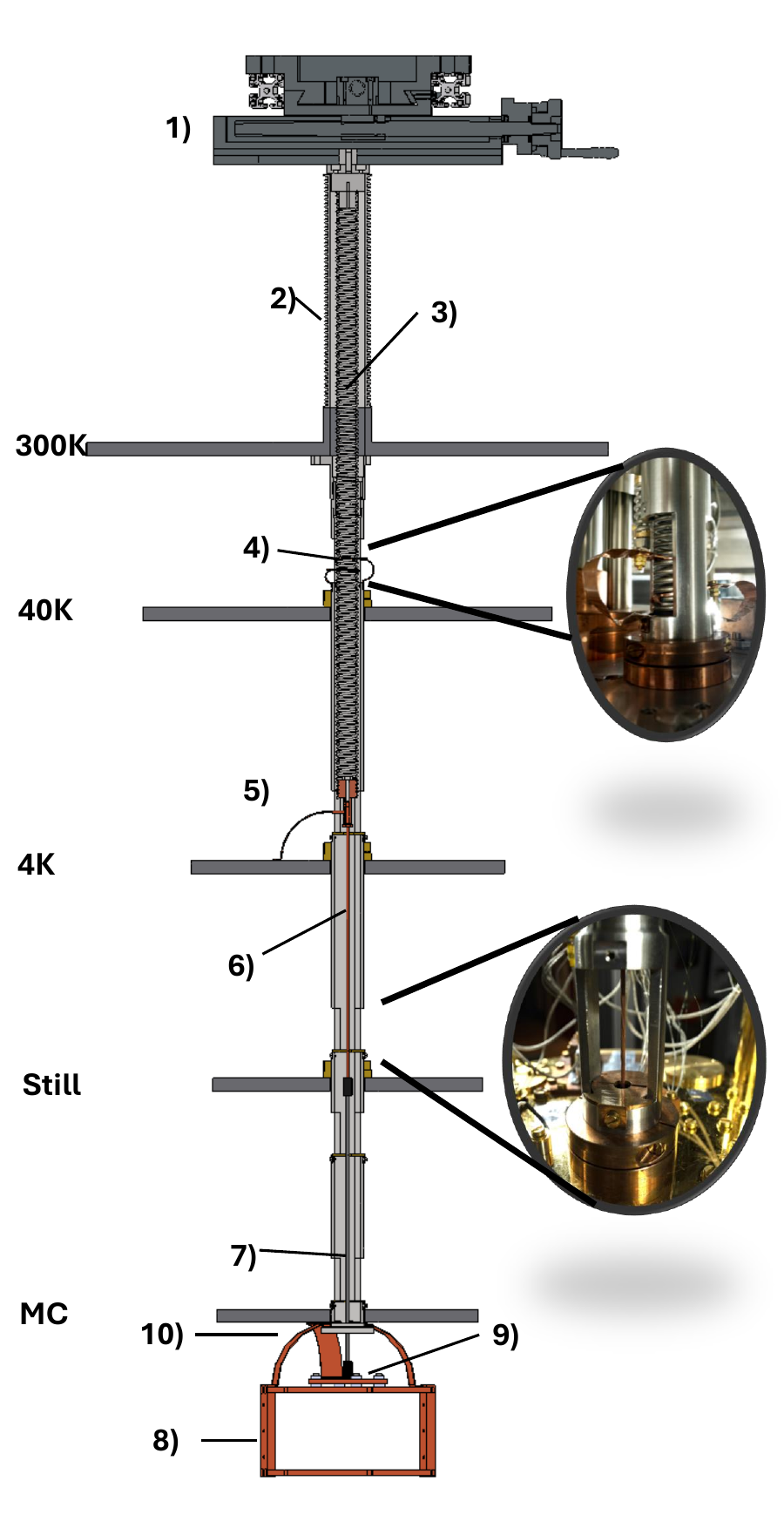}
      \caption{Schematic of the vibration decoupling system in a simplified cryostat for illustration, comprising in-plane positioning stage (1), vacuum bellow (2), steel extension spring (3), inner-spring radiation shields (4), 4\,K thermal link (5), copper wire (6), Kevlar string (7), detector box (8), thermal decoupling stage (9) and thermal links to mixing chamber (10). Details are described in the text. The inset pictures show the inner-spring radiation shields (top) and a piece of the copper section fed through a radiation shield at still level (bottom).}
      \label{fig:schematic}
   \end{figure}

   In order to avoid modifying the cryostat's basic components, the NUCLEUS vibration decoupling system focuses on isolating the target detectors from pulse tube and surrounding laboratory vibrations. This is primarily achieved by mounting the entire experimental volume on a frame of reference independent of the vibrating cryostat. By minimizing contacts between the two systems as much as possible, the propagation of pulse tube vibrations from the cryostat to the experimental volume can be effectively reduced. Additional passive attenuation of environmental vibrations is achieved by coupling the experimental volume to the new frame of reference via a 1.8\,m long spring-pendulum. This setup provides an easy and efficient way of damping vibrations, however creating a direct link between room temperature and the coldest stage of the cryostat. Special care is taken to achieve this temperature gradient spanning four orders of magnitude while minimizing the reintroduction of pulse tube vibrations into the experimental volume without influencing the cryostat's base temperature. \\
   A simplified schematic of the vibration decoupling system (VDS) is shown in figure \ref{fig:schematic}. Its main components are:
   \begin{enumerate}
       \item In-plane positioning stage with 50\,$\mu$m precision, fixed on an independent frame of reference. This can either be on the laboratory's ceiling or a dedicated standing rack. The VDS is mounted to this stage which is used for centering in order to avoid contacts between the spring-pendulum and the central rod.
       \item Hydroformed bellow to keep the vacuum between the cryostat's 300\,K plate and the positioning stage while keeping vibrational transfer low.
       \item Steel extension spring spanning from the positioning stage down towards the 4\,K stage of the cryostat. It hangs contact-free inside the cryostat, protected by a central steel rod.
       \item Radiation shields coupled to the 40\,K stage of the cryostat, protecting from 300\,K radiation transmitted through the inner part of the spring.
       \item Spring termination. The spring portion of the VDS ends here and is thermally linked to the 4\,K stage. 
       \item Copper wire extension. A copper wire thermalized at 4\,K is attached to the end of the spring and extends it further downwards.
       \item Kevlar insulation. A Kevlar string is connected to the copper wire and thermally decouples the detector box from the 4\,K temperature.
       \item Detector box holding the payload, which is connected to the Kevlar string.
       \item Thermal decoupling stage. A copper disk which is thermally insulated from the detector box (8) and at the same time coupled to the MC, thereby removing residual heat that is introduced through the Kevlar string or thermal radiation.
       \item Heat links. Flexible heat conducting bridges provide a thermal link between detector box (8) and MC, as well as decoupling stage (9) and MC.
   \end{enumerate}
   The extension spring (3) is terminated at the 4\,K stage for thermal reasons: Typical dry cryostats do not allow the use of thermal exchange gas. Therefore all components need to be cooled via direct thermal contact, which is not feasible over the entire length of a steel spring. However, the pulse tube driven 4\,K and 40\,K stage of the cryostat are robust against thermal radiation from an only partially cooled spring body due to their high cooling power of 2.0\,W (at 4.2\,K) and 55\,W (at 45\,K), respectively \cite{BF_PT420}. In contrast the lower stages need to be protected from any residual radiation. The spring is therefore continued with shorter, thin wire sections, on which thermal equilibrium is easier to reach. A stable temperature gradient below 4\,K can thus be realized. Radiation from hotter stages needs to be shielded effectively as shown in the inset pictures in figure \ref{fig:schematic}.

   \subsection{Vibration decoupling features}
   \label{sec:vibration_decoupling}
\begin{figure}[htbp]
    \centering
      \includegraphics[width=1\linewidth]{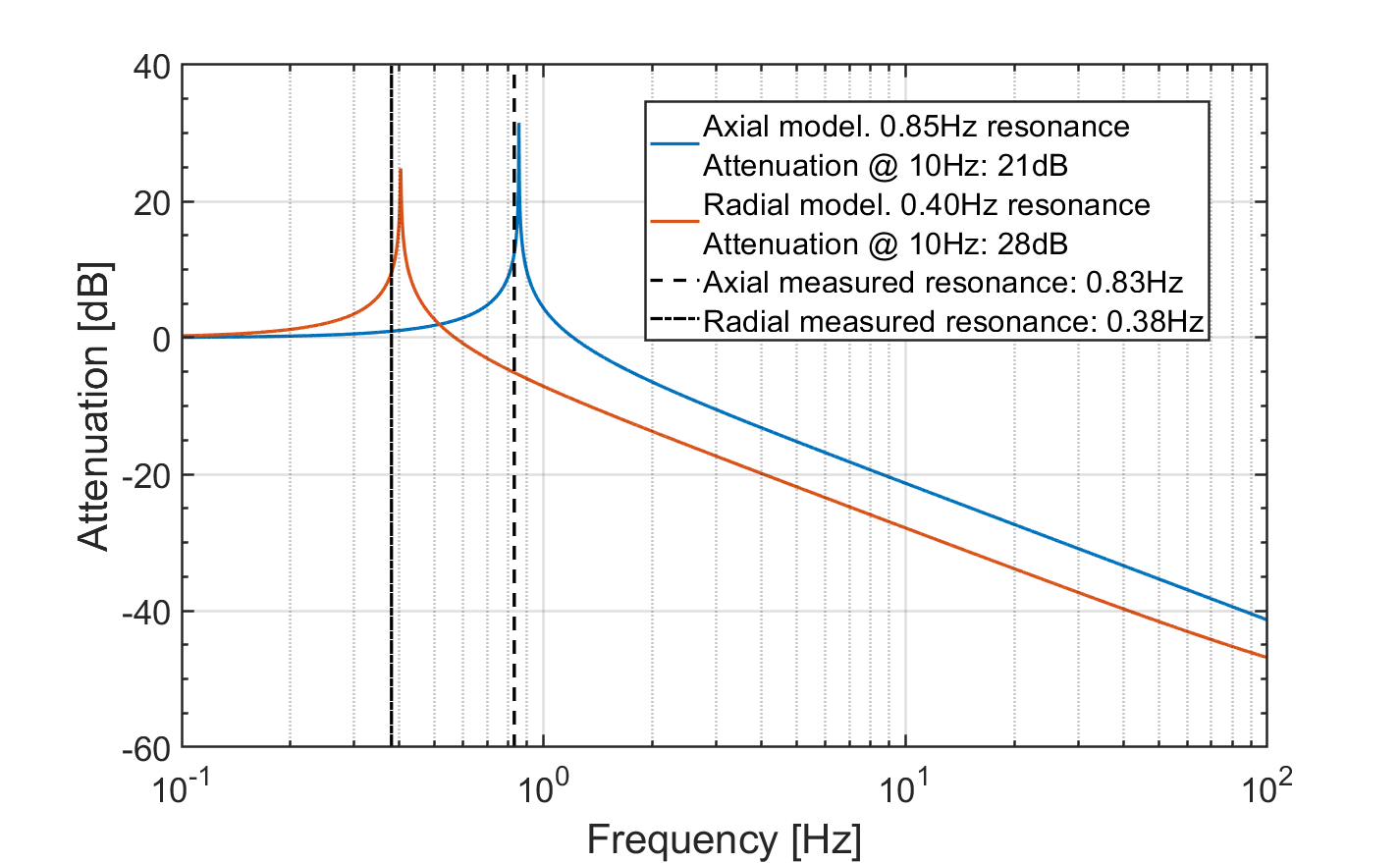}
      \caption{Theoretical transfer function of axial (blue) and radial (orange) component of the spring-pendulum used in the NUCLEUS experiment. The measured resonance frequencies of the systems lie within 5\,\% of the expected value. An attenuation of more than 20\,dB is predicted for both axial and radial direction above 10\,Hz.}
      \label{fig:transfer_function}
   \end{figure}

   While transmission of the majority of cryostat related vibrations is avoided by the independent mounting position of the pendulum, a limited amount is transmitted via the vacuum bellow and adds to remaining environmental vibrations. These are decoupled by the spring-pendulum itself. By taking the small-angle approximation, this can be described by a damped harmonic oscillator with independent equations of motion in axial (spring) and radial (pendulum) direction. As such, any motion with frequencies above $\sqrt{2}f_{0}$ is attenuated, with $f_{0}$ being the resonance frequency. In radial direction this is given by: 
   \begin{equation}
    \label{eq:pendulum_resonance}
       f_{0,radial} = \frac{1}{2\pi} \cdot \sqrt{\frac{g}{l}} \, ,
   \end{equation}
   where $l$ denotes the length of the pendulum. The pendulum length of 1.8\,m used in this work results in a radial resonance frequency of 0.40\,Hz. \\
   In axial direction the environmental vibrations are decoupled by the steel spring. The resonance frequency of which can be calculated using the mass $m \approx 7$\,kg of the detector box and the spring constant $k = 202$\,N/m:
   \begin{equation}
    \label{eq:spring_resonance}
        f_{0,axial} = \frac{1}{2\pi}\cdot\sqrt{\frac{k}{m}} \; ,
   \end{equation}
   yielding an axial resonance frequency of 0.85\,Hz for the presented setup. \\
   The expected attenuation of environmental vibrations by the spring-pendulum in axial and radial direction can be estimated using the displacement transmissibility $|H(\omega)|$ given by \cite{mori2017mechanical}:
   \begin{equation}
    \label{eq:transfer_function}
       |H(\omega)| = \frac{1}{\sqrt{\left[1-(\frac{\omega}{\omega_0})^2\right]^2 + (2\zeta\frac{\omega}{\omega_0})^2}} \; ,
   \end{equation}
   with $\omega_0$ being the undamped resonance frequency and $\zeta$ the reduced damping coefficient ($\zeta = 1$ corresponding to critical damping). The result is shown in figure \ref{fig:transfer_function}.   
  The actual resonance frequencies at room temperature were found to be 0.83\,Hz (0.38\,Hz) in axial (radial) direction using accelerometer measurements performed on the uncoupled spring-pendulum. These results fit the expected values within 5\,\% and are indicated in the figure as vertical black lines. For the achieved resonance frequencies an attenuation of environmental vibrations well above 20\,dB is expected in the region of interest above 15\,Hz, assuming the benchmark case with negligible influence of the thermal connections between croystat and VDS. Depending on size and stiffness, these may have a visible effect on the VDS performance, as discussed in section \ref{sec:experimental_results}. Ideally the stiffness of all these links is well below the stiffness of the spring-pendulum.\\
 
    \subsection{Mitigation of Heat Sources}
    \label{sec:thermalization}
\begin{figure}[htbp]
    \centering
      \includegraphics[width=1\linewidth]{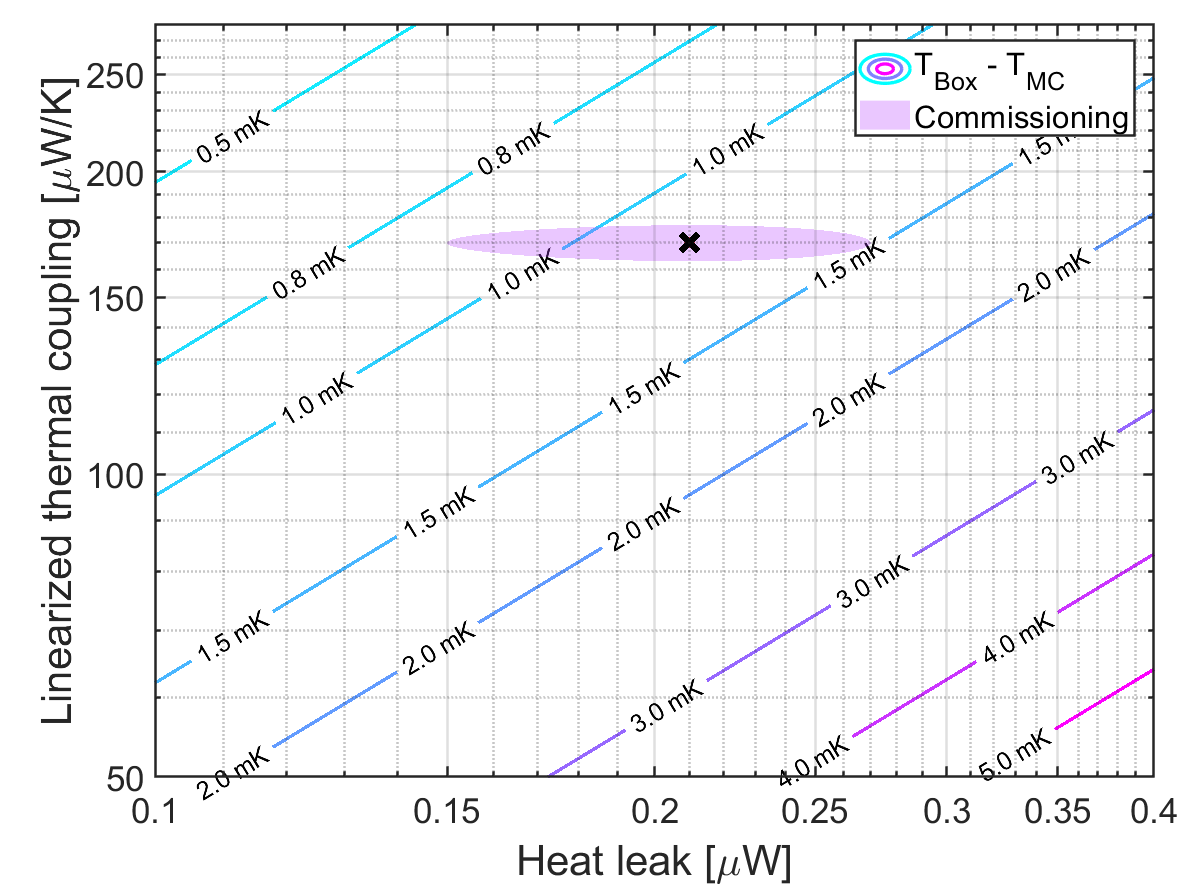}
      \caption{Resulting temperature differences between detector box and cryostat mixing chamber for different values of induced heat and thermal coupling strength linearized at 8\,mK. The contour shows the 1\,$\sigma$ uncertainty region obtained during the VDS commissioning run (see section \ref{sec:experimental_results}).}
      \label{fig:heat_leak}
   \end{figure}

   The pendulum needs to maintain a stable temperature gradient over more than four orders of magnitude from room temperature down to less than 10\,mK without negatively influencing the cryostat's base temperature. Identification and mitigation of possible heat sources is key.\\
   Two different heat load budgets need to be considered:
   \begin{enumerate}
       \item The detector box temperature depends on the injected heat load and the strength of its thermal link to the mixing chamber. As the latter is also limited by the stiffness constraints, even small heat loads on the detector box can lead to a large temperature difference to the mixing chamber. Thermal link design is therefore a trade-off between VDS performance and achieved heat load.
       \item The total heat load on the mixing chamber must be sufficiently low to keep the mixing chamber close to base temperature. For the detector box's target temperature $\leq$ 10\,mK a mixing chamber temperature below 8\,mK is preferred, as higher mixing chamber temperatures would put too high demands on the thermal coupling between detector box and mixing chamber. To maintain this mixing chamber temperature, the total additional heat load must stay below 1\,$\mu$W.
   \end{enumerate}
   The required thermal link for a given heat load on the detector box can be estimated using Fourier's law of heat conduction. For the case of metallic links at low temperatures the thermal conductivity $k$ scales linearly with temperature \cite{ventura2010art}: $k = a\cdot T$ with proportionality constant $a$, which leads to a quadratic dependency of the heat leak on temperature:
    \begin{equation}
    \label{eq:fourier_heat_int}
        Q = \frac{1}{2} \cdot g \cdot (T_{Box}^2 - T_{MC}^2) \, .
    \end{equation}
    with $T_{Box}$ and $T_{MC}$ being the temperatures of the detector box and mixing chamber, respectively, connected via a thermal link of conductivity $g$. $g$ is defined by the material's thermal conductivity proportionality constant $a$ and the link's geometry given by cross section area $A$ and length $L$ as $g = a\cdot\frac{A}{L}$. Using $T_{MC} \approx 8$\,mK, equation \ref{eq:fourier_heat_int} is visualized in figure \ref{fig:heat_leak}, where $g$ is linearized at $T_{MC}$ to yield a value in the commonly used units of $\mu W/K$.
   A highly conductive thermal link with low stiffness is required. Woven copper braids provide a commonly available solution. We chose braids with a flat geometry consisting of 24$\times$109$\times$0.07\,mm copper strands, amounting to a cross section of 10\,mm$^2$. The conductivity of copper at low temperatures depends linearly on the residual-resistance ratio (RRR), a measure for the material's purity \cite{ventura2010art}. For the copper used it was measured with a liquid helium dipstick to be $(78 \pm 3)$. For a single braid of 20\,cm length we therefore expect a thermal coupling of $(61 \pm 8)$\,$\mu$W/K at 10\,mK, amounting to a total coupling of $(182 \pm 22)$\,$\mu$W/K when using three braids between detector box and mixing chamber. This in turn limits the heat leak budget on the detector box to a few hundreds of nano-watts. As the purity of the copper used is only average, investigations are currently ongoing to strongly improve the thermal coupling using high purity copper and annealing techniques. \\
   In the following the different identified sources of residual heat and the efforts taken to keep them below the allowed heat load budgets are discussed.

    \subsubsection{Thermal contact}
    \label{sec:thermal_contact}
    The vibration decoupling system creates a direct connection between 10\,mK and room temperature. In order to maintain this huge temperature gradient without exceeding the heat leak budget, thermalization of the spring-pendulum happens over multiple stages, as also indicated in figure \ref{fig:schematic}. Thermal anchoring at 4\,K happens via a flexible copper cable (5,6), which is effectively decoupled from 300\,K by the spring (3) itself. Its total wire length of around 13\,m combined with the low thermal conductivity of steel alloys \cite{bradley2013properties} results in a heat load on the 4\,K stage approaching 2\,mW as the spring cools down, which leads to no significant temperature increase of the stage. The decoupling between 4\,K temperature and the detector box (8), which is thermally anchored at 10\,mK, is realized with a Kevlar string (7), due to its high durability at small diameters paired with minimal thermal conductivity at low temperatures \cite{bradley2013properties}. A Kevlar wire with 1\,mm thickness and 30\,cm length leads to a sufficiently low heat leak below 100\,nW. \\
    As an extra safety measure the Kevlar is connected to a copper plate (9) that is thermally coupled to the mixing chamber, but decoupled from the detector box using thick plastic washers. By redirecting the excess energy directly to the mixing chamber, this reduces remaining heat leaks from the Kevlar and residual radiation at the cost of an additional thermal link. 
    
   \subsubsection{Thermal radiation}
   \label{sec:radiation}
   Thermal radiation is a topic of increasing importance when approaching lower temperatures. While the available cooling power generally decreases exponentially with temperature, the power emitted by thermal radiation is scaling with $T^4$ as described by the Stefan-\\Boltzmann-Law:
   \begin{equation}
       \label{eq:stefan-boltzmann}
       Q = \sigma A \epsilon T^4
   \end{equation}
   $A$ denotes the surface area of the radiating body, $\epsilon$ its emissivity and the Stefan-Boltzmann constant \\$\sigma = 5.67 \times 10^{-8}$W/(m$^2\cdot$T$^4$). Polished metal surfaces like gold or copper can have emissivity values below 0.1 \cite{pobell2007matter}, however in order to give conservative estimates we assume $\epsilon = 1$ in the following.\\
   Shielding radiation is a delicate balance between avoiding direct line of sight through the central rod while keeping the pendulum from touching the cryostat. Using equation \ref{eq:stefan-boltzmann} one can estimate the total area of direct line of sight that is permitted from each cryostat stage to the detector box for a heat budget of hundreds of nano-watts. For radiation originating from the still stage several square meters of face area are tolerable, 100\,cm$^2$ from the 4\,K will already have a limiting effect. Radiation from higher stages needs to be shielded effectively, as from the 40\,K stage 1\,mm$^2$ uses up the entire heat budget and radiation from room temperature is critical from several 100\,$\mu$m$^2$ on, meaning every macroscopic hole would lead to an unacceptably high temperature on the detector box.\\
   It is thus imperative to have the predominant part of the area facing the detector box at or below still temperature. In order to close the direct line of sight to the detector box from room temperature introduced by the central 
   rod and spring pendulum, several measures are taken: \\
   (i) A copper threaded plug is located at the spring's endpoint and thermally coupled to the 4\,K stage, thus blocking radiation transmitted through the spring's inner diameter. \\
   (ii) In order to achieve a temperature close to 4\,K on this piece, the 300\,K radiation is initially shielded by inserting two radiation shields into the spring at the 40\,K stage (see the top inset picture in figure \ref{fig:schematic}). By using two shielding layers, a thermal coupling of 3\,mW/K to each shield is sufficient to have the lower one at a temperature of 40\,K by the same principle as used in multi-layer insulation (MLI). At that temperature this can be achieved using thin and flexible copper bands which do not significantly transmit cryostat vibrations to the spring system.\\
   (iii) Tailored radiation shields with a 4\,mm diameter central hole to allow the lower parts of the vibration decoupling system to pass are installed at every lower stage starting from 4\,K to block radiation transmitted through the gap between spring and central rod (see bottom picture in figure \ref{fig:schematic}). While the first of these shields is located at 4\,K temperature, the lower ones are at still temperature and therefore shine a negligible amount of radiation onto the detector box, with the last one located at the end of the central rod. Having multiple stages of these partially open radiation shields increases the number of reflections required for secondary radiation to transmit to the detector box, leaving only the contribution from the 4\,K threaded plug in direct line of sight to the detector box. However, given the central hole diameters of 4\,mm, this amounts to less than 1\,nW and thus is also negligible.\\
   (iv) The thermal decoupling stage mounted on top of the detector box acts as a final radiation shield thermally anchored to the mixing chamber stage, to reduce any residual radiation that is reflected towards the detector box.
   
   \subsubsection{Energy dissipation via damping}
   \label{sec:damping_dissipation}
   Energy injected into the VDS is stored in the spring-pendulum and creates a relative motion compared to the cryostat. The unavoidable connections between cryostat and VDS, namely thermal links and any type of read-out cable, dampen this motion, thereby dissipating the stored energy into heat. While this effect is completely negligible for higher stages of the cryostat, it may have a visible impact on the temperature reached on the detector box, depending on the thermal coupling and amount of damping. Measurements have shown that the dominating contribution in the presented setup can be attributed to the thermal links between mixing chamber and detector box, decreasing the damping time from several tens of seconds to typical values of around 3\,s. In the vibrational sense, extra damping has the downside of decreasing attenuation at high frequencies, but at the same time increasing run-time stability by decreasing the time it takes for the system to stop moving after being excited by an external shock. In the following we give an estimate about the expected thermal impact of strong damping.\\
   The viscous damping force acting on the system is given by: 
   \begin{equation}
    \label{eq:damping_force}
       F = \gamma \cdot \Delta \dot X \;.
   \end{equation}
   $\Delta \dot X$ denotes the relative velocity of the two stages, i.e. the mixing chamber and detector box, which have a mechanical link with damping constant $\gamma = \frac{m}{\tau}$ with detector box mass $m$ and damping time constant $\tau$.
   The expected energy dissipation per second, $P$, can then be calculated as:
   \begin{equation}
    \label{eq:damping_dissipation}
       P = F \cdot \frac{\Delta X}{\Delta t} = \gamma \cdot |\dot X - \dot X_0|^2 \; .
   \end{equation}
   $\dot X$ and $\dot X_0$ are the velocities of the mixing chamber and the detector box, respectively. \\
   In Fourier space we can disentangle the contribution from each vibrational mode. Making use of the properties of the Fourier transform, equation \ref{eq:damping_dissipation} can be rewritten in a form taking the more commonly used displacement spectra as an input and assuming the phases of $X$ and $X_0$ to be uncorrelated:
   \begin{equation}
    \label{eq:damping_dissipation_fourier}
       P(\omega) = \gamma \cdot \omega^2 \cdot |X(\omega)^2 + X_0(\omega)^2|
   \end{equation}
   For $X$ and $X_0$ we insert accelerometer spectra obtained at the mixing chamber and detector box level with all thermal links and read-out cables installed. The damping time constant of this setup was measured at warm to be $\tau \approx 3$\,s. The extreme cases of full phase correlation ($|X(\omega) + X_0(\omega)|^2$) and anti-correlation ($|X(\omega) - X_0(\omega)|^2$) are used as lower and upper limits.  In total a dissipated power of $(7.07 \pm 0.64)$\,nW is expected for axial movements, $(43.9 \pm 9.2)$\,nW in radial direction. Only half of that is transmitted onto the detector box, the other half directly flows to the mixing chamber. We thus expect energy dissipation via damping to be a second order effect in this case.
   
     \section{Experimental Results} 
\label{sec:experimental_results}
The vibration decoupling system has been in use for more than 30 cooldowns in two different LD400 cryostats located in an above- and a shallow underground laboratory, respectively. While for technical reasons the thermal and vibrational behaviour of the VDS was characterized during a dedicated commissioning run, presented results for detector operation with the VDS were obtained during two physics runs spanning several weeks, namely an XRF calibration and a background characterization. 
\subsection{Cooldown Time and Base Temperature}
\label{sec:cooldown_data}
 \begin{figure}[htbp]
    \centering
      \includegraphics[width=1\linewidth]{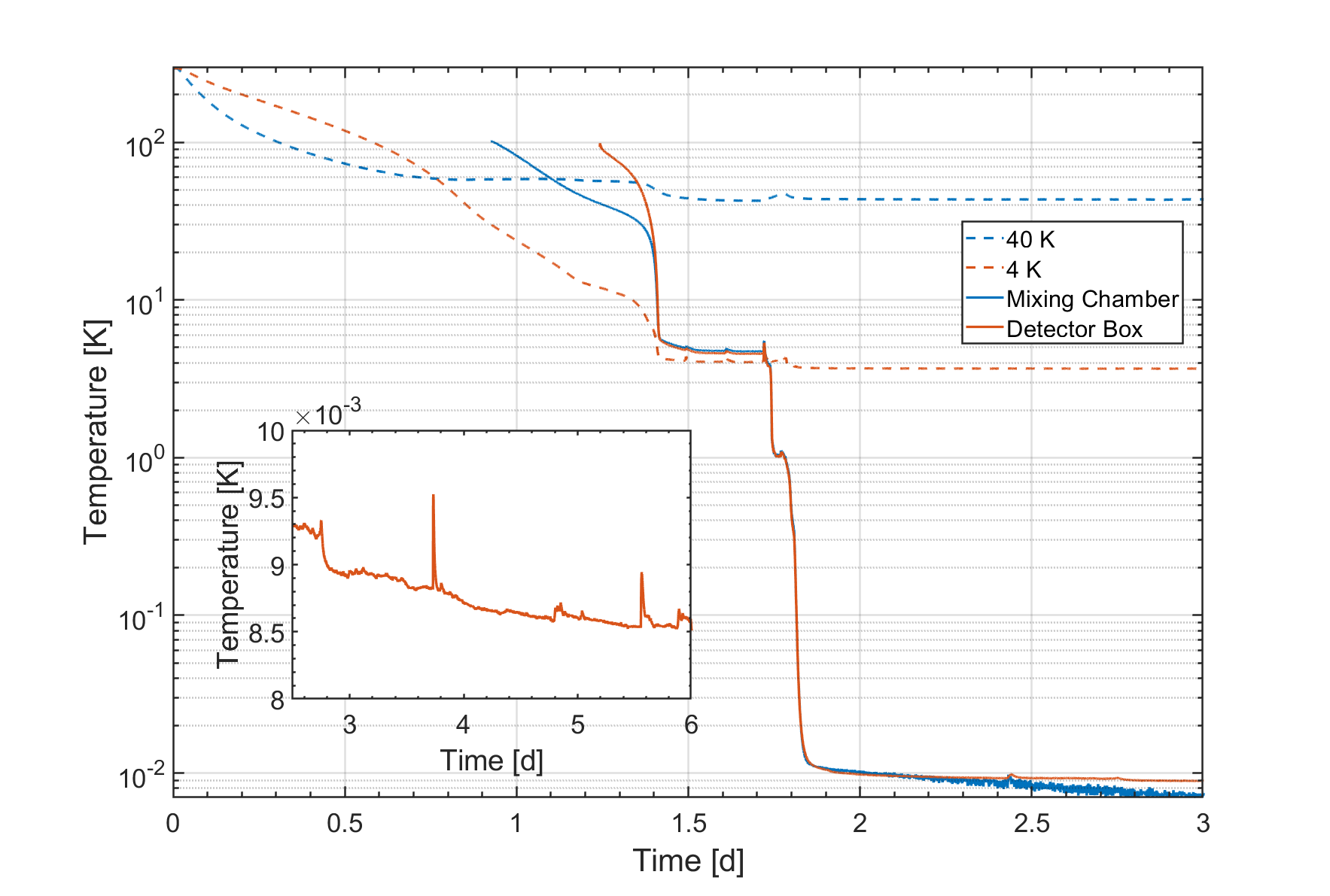}
      \caption{Cooldown curves from NUCLEUS VDS commissioning run with mixing chamber (solid blue) and detector box temperature (solid orange) both reaching temperatures below 9\,mK after less than three days of total cooldown time. Temperature stability on the detector box over a period of several days is shown on the inset. The dashed lines show the 4\,K (orange) and 40\,K (blue) stage temperatures for reference.}
      \label{fig:Cooldown_curve}
   \end{figure}
In the VDS commissioning run a temperature around 10\,mK on the suspended detector box could be reached after less than 2 days of cooling time and a stable temperature below 9\,mK was achieved after less than 3 days, as shown in figure \ref{fig:Cooldown_curve}. The temperature on the detector box remained stable within less than 1\,mK, as is shown on the inset. According to the manufacturer's specifications, an empty LD400 is expected to reach a base temperature of $<$8\,mK after approximately one day of cooling time \cite{BF_LD}. Correspondingly the VDS increases the cooling time by less than a day and increases the base temperature by approximately 1\,mK.
The detector box was coupled strongly to the mixing chamber using three copper braids (as described in section \ref{sec:thermalization}). Using a 100\,$\Omega$ heater on the detector box, the thermal coupling between box and mixing chamber was measured to be (170 $\pm$ 7)\,$\mu$W/K which fits the theoretical prediction (see section \ref{sec:thermalization}) within uncertainty and gives a cooling power around around 2.5\,$\mu$W at 20\,mK. Around 15\,\% of the cooling power available at the mixing chamber stage is preserved on the detector box, which is subject to a residual heat leak of (0.21 $\pm$ 0.06)\,$\mu$W. Turning off the pulse tube for a short duration does not lead to significant temperature changes on the detector box, giving an upper limit to vibration-induced dissipation through damping in the thermal links of 50\,nW, consistent with the values predicted in section \ref{sec:damping_dissipation}. The resulting values for the heat leak and thermal coupling are also marked in figure \ref{fig:heat_leak}.\\

\subsection{Vibration Measurements}
\label{sec:vib_measurements}
 \begin{figure}[h]
    \centering
      \includegraphics[width=1\linewidth]{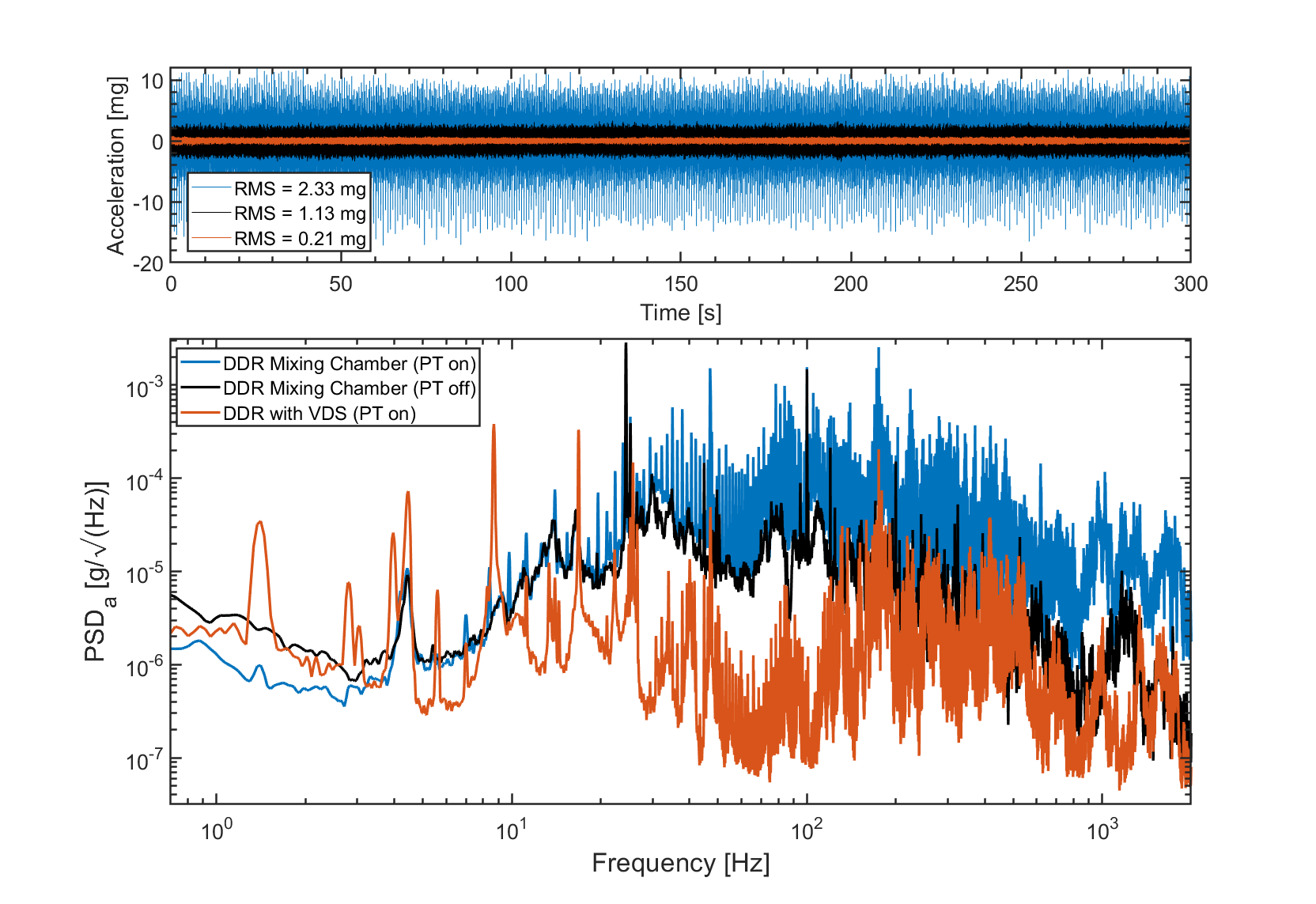}
       \includegraphics[width=1\linewidth]{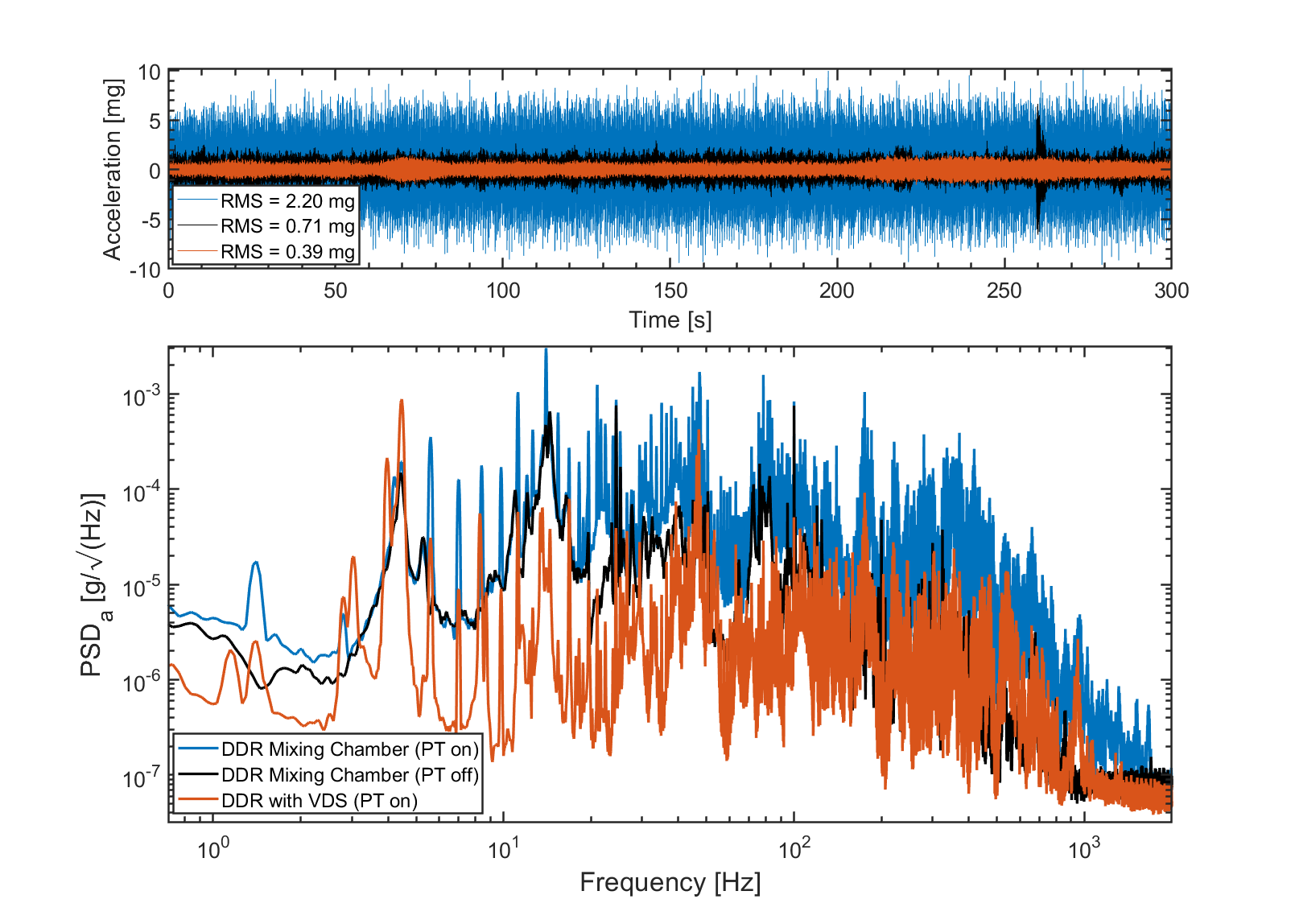}
      \caption{Axial (top) and radial (bottom) acceleration spectra taken on the DDR MC with PT on (blue), PT off (black) and on the vibration decoupling systems' detector box with PT on (orange). The decoupling system shifts most of the motion into its eigenmodes at lower frequencies. The total RMS on the detector box with continuously operating pulse tube lies well below the total RMS on the mixing chamber stage with pulse tube off.}
      \label{fig:decoupled_spectra_axial}
   \end{figure}
Using the vibration measurement setup described in section \ref{sec:equipment}, we quantified the performance of the vibration decoupling system after the VDS comissioning run by comparing vibration spectra measured on the detector box with those taken on the cryostat's mixing chamber. The corresponding acceleration spectra are shown in figure \ref{fig:decoupled_spectra_axial}. The time domain data shows vibration levels obtained on the detector box which are about an order of magnitude lower than on the mixing chamber stage, when the pulse tube is running. The vibration level on the detector box with operating pulse tube is even lower than the one on the mixing chamber without pulse tube operation. This demonstrates a vibration level obtained during continuous cryostat operation below typical environmental vibrations. \\
Acceleration and displacement RMS values computed for the frequency ranges specified in section \ref{sec:equipment} are displayed in table \ref{tab:axial_result}. The effect of the spring based pendulum becomes directly apparent when comparing the different frequency regions. Most injected energy is stored in the system's radial and axial eigenmodes. Attenuation starts around 10\,Hz in axial direction (5\,Hz in radial direction) and remaining motion in the region of interest is of nm-scale in axial direction (sub-nm-scale in radial direction). This corresponds to more than a factor 50 reduction in axial direction compared to the mixing chamber stage with pulse tube running and still almost a factor 30 reduction compared to the case without pulse tube. \\
\begin{table*}[htbp]
    \centering
    \begin{tabular}{|c|c|c|c|c|}
    \hline
     \multicolumn{5}{|c|}{Axial}\\
      \hline
      Frequency range &  VDS PT off& VDS PT on & Attenuation & Attenuation\\
      $ $[Hz] & [nm] & [nm] & MC PT on/VDS PT on & MC PT off/VDS PT on\\
       \hline
        15-1000 & 1.89 $\pm$ 0.16 & 3.06 $\pm$ 0.26  & 56.8 $\pm$ 9.7 & 28.6 $\pm$ 4.9 \\ 
       \hline
         &  [$\mu$g] & [$\mu$g] & &  \\
       \hline
        0-15 & 135$\pm$ 12 & 126 $\pm$ 11 & 0.373 $\pm$ 0.064 & 0.454 $\pm$ 0.079 \\ 
        15-1000 & 20.0 $\pm$ 1.7 & 80.9 $\pm$ 6.9  & 27.6 $\pm$ 4.7 & 13.2 $\pm$ 2.2  \\ 
        0-2000 & 155 $\pm$ 13 & 208 $\pm$ 18 & 11.1 $\pm$ 1.9 & 5.41 $\pm$ 0.92   \\
       \hline  
        \hline
     \multicolumn{5}{|c|}{Radial}\\
      \hline
       Frequency &  VDS PT off& VDS PT on & Attenuation & Attenuation\\
       range [Hz] & [nm] & [nm] & MC PT on/VDS PT on & MC PT off/VDS PT on\\
       \hline
        15-1000 & 0.153 $\pm$ 0.013 & 0.171 $\pm$ 0.015 & 225 $\pm$ 38 & 44.4 $\pm$ 7.6  \\ 
        \hline
         &  [$\mu$g] & [$\mu$g]  & & \\
       \hline
        0-15 & 139 $\pm$ 12 & 306 $\pm$ 26 & 3.62 $\pm$ 0.62 & 1.48 $\pm$ 0.25 \\ 
        15-1000 & 5.60 $\pm$ 0.48 & 87.5 $\pm$ 7.4 & 12.5 $\pm$ 2.1 & 2.48 $\pm$ 0.42 \\ 
        0-2000 & 144 $\pm$ 12 & 393 $\pm$ 33 & 5.60 $\pm$ 0.95 & 1.70 $\pm$ 0.29 \\
       \hline
    \end{tabular}
    \caption{RMS values obtained in vibration measurements on the NUCLEUS VDS for different frequency regions. The attenuation of the vibration decoupling system with pulse tube running compared to the cryostat's mixing chamber with and without pulse tube operation are shown. }
    \label{tab:axial_result}
\end{table*}
Radial motion is reduced by more than two orders of magnitude in the region of interest compared to the mixing chamber stage with pulse tube running, and more than a factor of 40 without pulse tube. These values prove that we do not only manage to decouple pulse tube vibrations effectively, but also strongly decrease environmental vibrations that are propagated into the cryostat. Remaining vibrations are mostly focused in the frequency region below the ROI, in which radial decoupling is weak and axial vibrations are even slightly enhanced. This effect and the higher attenuation frequency compared to the expected value from the spring-pendulum's resonance frequencies are due to the thermal links between cryostat and VDS. The flat profile of the copper braids used as thermal link in this work also leads to an overall larger attenuation achieved in axial direction. By using thermal links with a round profile instead, the attenuation in both directions could be more evenly distributed.\\
Two additional sets of measurements were performed to quantify their impact, which are summarized in table \ref{tab:coupling_effect}. 
 \begin{table}[htbp]
    \centering
    \begin{tabular}{|c|c|c|}
    \hline
      VDS to cryostat & Axial RMS & Radial RMS \\
       couplings & [mg] & [mg] \\
       \hline
       Full & 0.21 & 0.39 \\
       \hline
       Reduced & 0.18 & 0.10 \\
       \hline
       None & 0.17 & 0.17\\
       \hline
    \end{tabular}
    \caption{Effect of the different couplings between cryostat and VDS on the measured acceleration RMS. "Full" refers to the case shown in figure \ref{fig:decoupled_spectra_axial}, "Reduced" to the setup without thermal links to the mixing chamber and for "None" all thermal couplings, radiation shields and read-out cables were removed. The dominating impact on the RMS is due to the thermal links to the MC stage.}
    \label{tab:coupling_effect}
\end{table}
Removing only the thermal links to the MC stage yields decreased acceleration RMS values in axial direction of 0.18\,mg (14\,\% decrease) and 0.10\,mg in radial direction (74\,\% decrease). Removing all remaining couplings except for the vacuum bellow only has a limited impact on the overall acceleration RMS but decreases the cut-off frequency in both radial and axial direction below 4\,Hz. These values indicate the potential of a significant VDS performance gain by developing thermal links with higher flexibility at comparable coupling strength, efforts on which are currently ongoing. Assuming ideal thermal links the remaining cut-off frequency is limited by the vacuum bellow and environmental vibration levels. The bellow's stiffness could be reduced further by increasing the distance between cryostat top plate and VDS reference frame, environmental vibration levels by further decoupling the reference frame at room temperature. 

\subsection{Cryogenic Detector Operation}
\label{sec:detector_operation}
\begin{figure}[htb]
    \centering
      \includegraphics[width=1\linewidth]{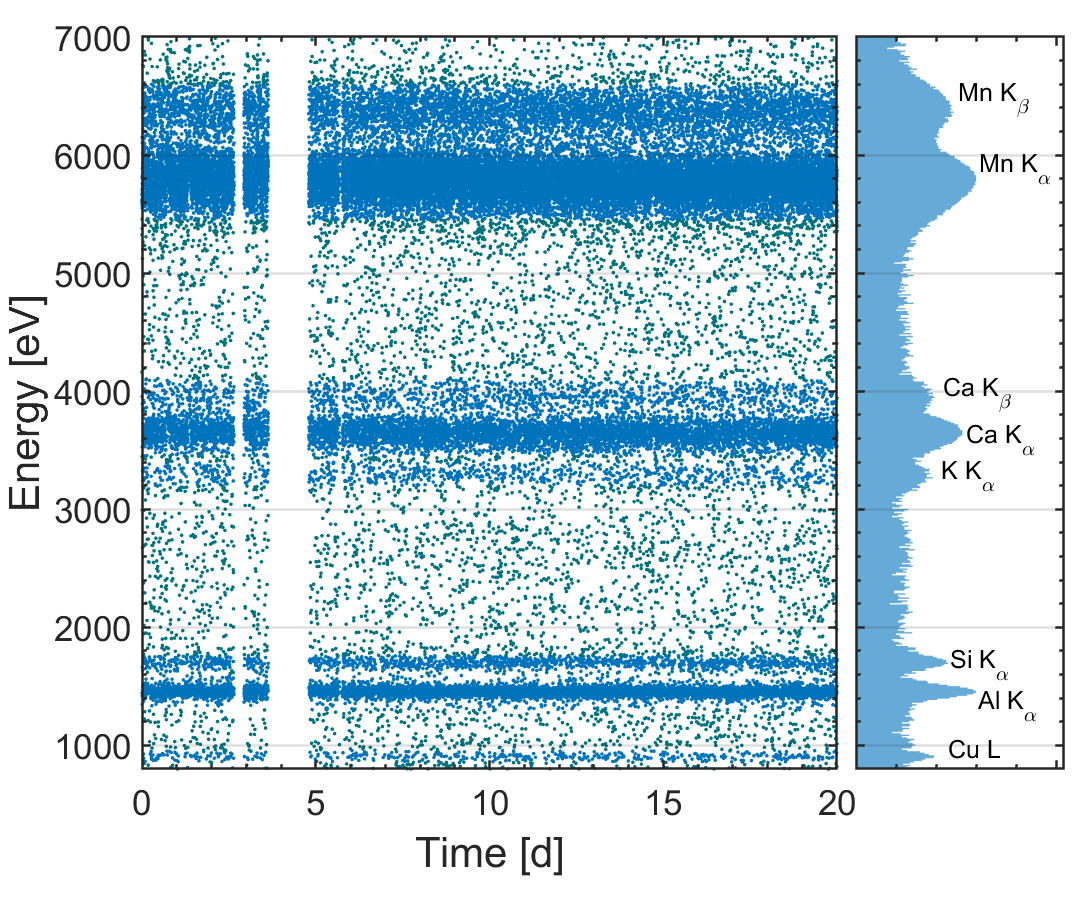}
      \caption{Stability plot of continuous detector data taking over a period of almost 3 weeks during an XRF calibration measurement. For details on the conducted XRF measurement and detector performance please refer to reference \cite{XRF}.}
      \label{fig:stability}
\end{figure}

\begin{figure}[htb]
    \centering
      \includegraphics[width=1\linewidth]{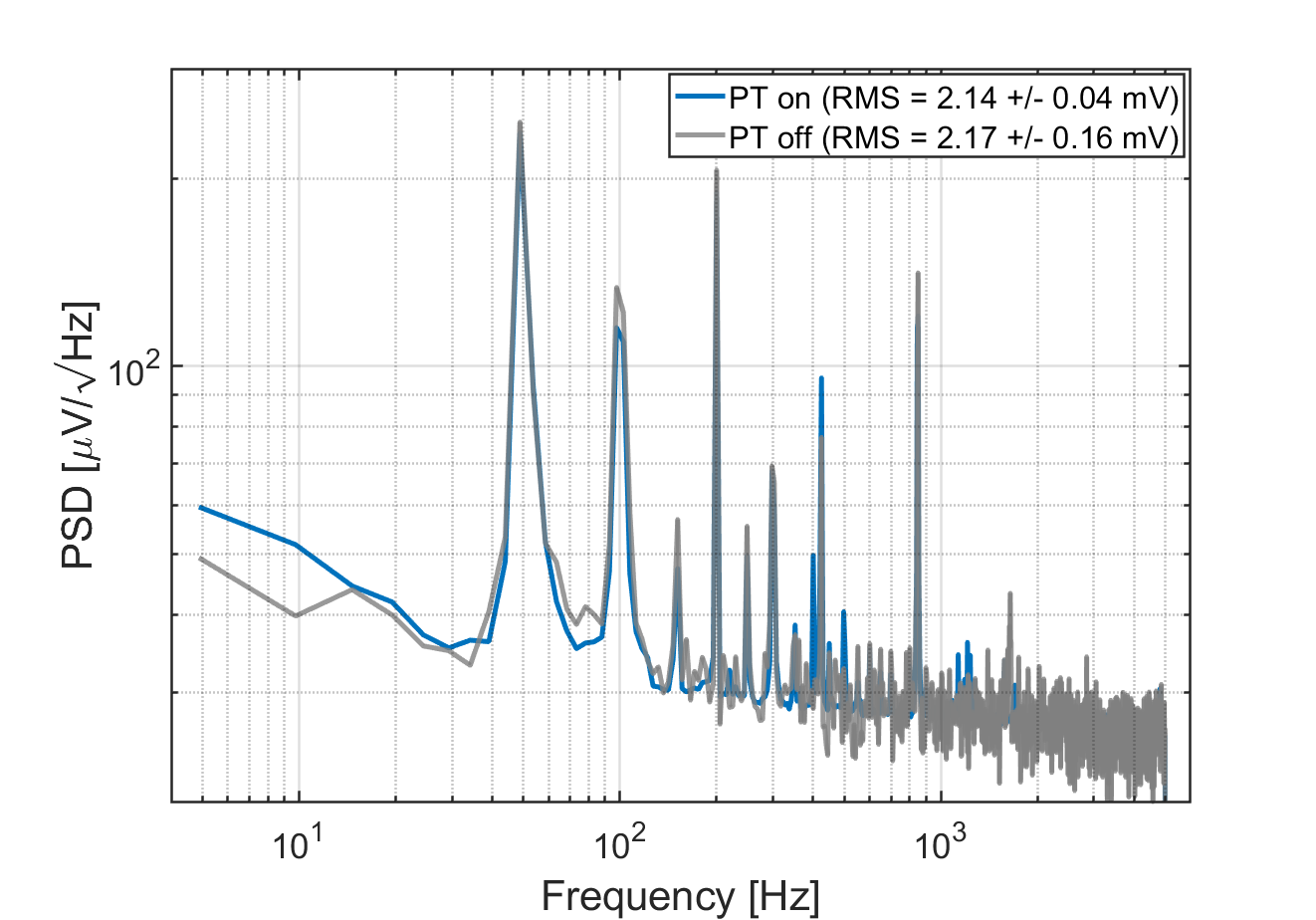}
      \caption{NPS comparison of a CaWO$_4$ target cube during the NUCLEUS background measurement for pulse tube switched on and off. Corresponding noise RMS values do not show any significant difference.}
      \label{fig:detector_operation}
\end{figure}

As the NUCLEUS experiment aims to detect rare events from reactor anti-neutrino CE$\nu$NS interactions at recoil energies mainly below 100\,eV, high stability together with ultra-low recoil energy thresholds are indispensable. This requires optimum detector performance with operating pulse tube at a stable base temperature below 10\,mK. Using the vibration decoupling system presented, cryogenic detectors can be operated continuously below 10\,eV baseline resolution, while the operation of the same detectors directly on the mixing chamber was not possible due to the high vibration level caused by the pulse tube. This value matches state-of-the-art performance of similar TES based calorimeters in wet cryostats \cite{rothe2021low}. To illustrate performance and stability, we exemplarily present results from two different data taking runs during the commissioning phase of the NUCLEUS experiment. In early 2023 an XRF calibration measurement was conducted \cite{XRF}. The setup consisted of a $^{55}$Fe source that irradiated a two-stage target arrangement. This causes emission of calibration lines from 677\,eV to 6.5\,keV. The source was used to calibrate a 0.75\,g CaWO$_4$ crystal. With the entire assembly mounted on the vibration decoupling system presented in this work, a stable detector operation with running pulse tube was obtained over 20 days, as shown in figure \ref{fig:stability}.
This emphasizes the advantage over the use of wet DRs, requiring a refilling of cryo-liquids every few days. Further details on this measurement are presented in \cite{XRF}.\\
\begin{figure}[htb]
    \centering
      \includegraphics[width=1\linewidth]{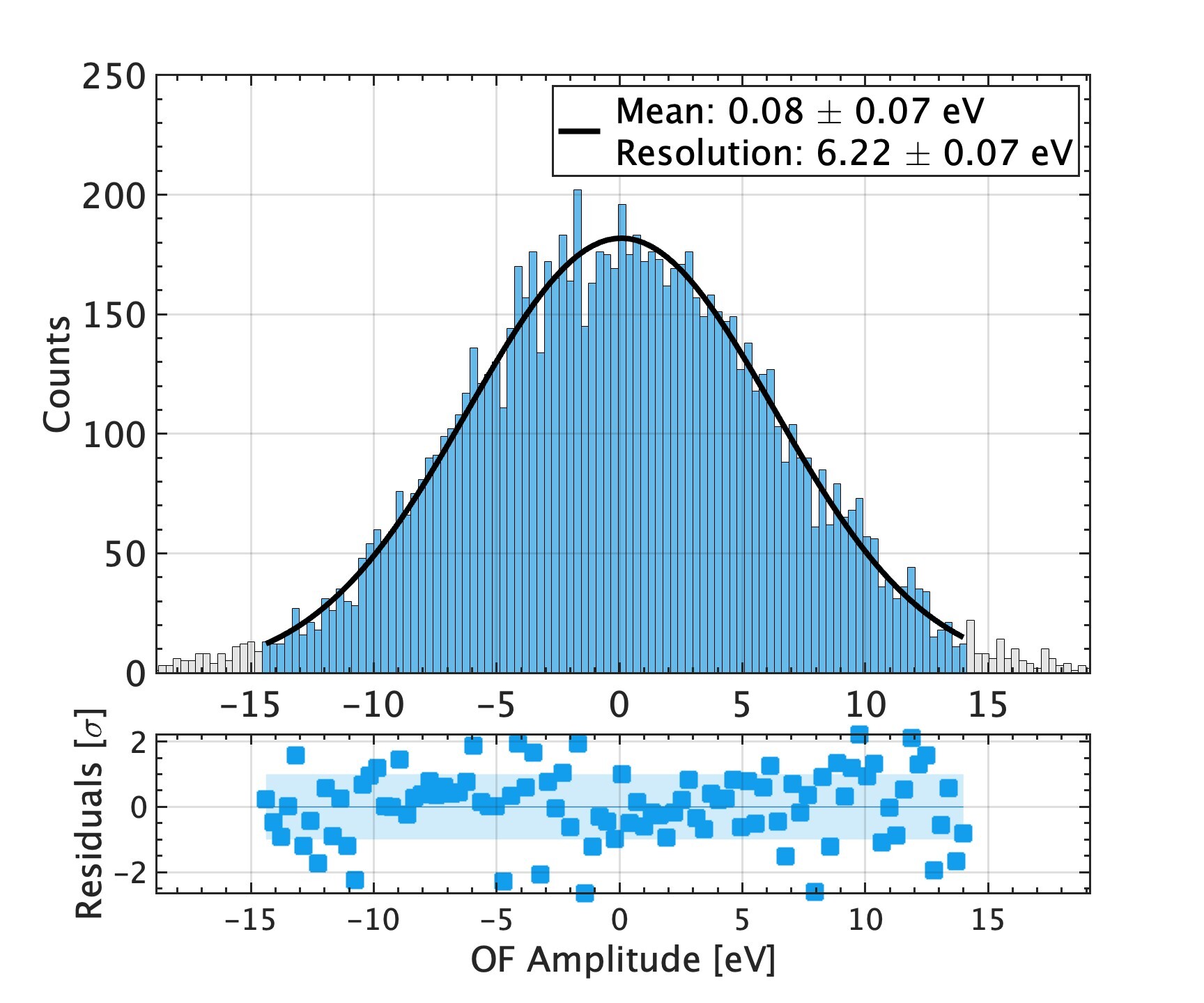}
      \caption{Baseline resolution of $(6.22 \pm 0.07)$\,eV obtained using optimal filtering (OF) on a gram-scale CaWO$_4$ cryogenic calorimeter that was operated on the VDS during a background measurement in the scope of the NUCLEUS experiment.}
      \label{fig:BR}
\end{figure}   
During an 8 weeks background measurement in the summer of 2024 a detector resolution of $(6.22 \pm 0.07)$\,eV after optimal-filtering was achieved in an 8 hour measurement on a gram-scale CaWO$_4$ calorimeter during continuous pulse tube operation (see figure \ref{fig:BR}). To check for the presence of vibration-related noise, a short file with the pulse tube turned off was taken. A comparison of the corresponding NPS obtained is shown in figure \ref{fig:detector_operation}. No significant differences are observed, proving that the detector's baseline resolution is not limited by vibrations. Remaining peaks in the noise power spectrum (NPS) at 50\,Hz and related harmonic frequencies were identified to be of electrical origin. \\

    
	\section{Conclusion and Outlook} 
	\label{sec:conclusion}
    In this paper we have described the development of a vibration decoupling system for dry dilution refrigerators in the scope of the NUCLEUS experiment, with the aim of achieving continuous cryogenic detector operation undisturbed by both environmental and pulse tube related vibrations. It is based on the two main approaches of mounting the experimental volume on a frame of reference independent of the cryostat and decoupling remaining vibrations using a 1.8\,m long spring pendulum. The resulting direct connection between room temperature and millikelvin temperatures poses a demanding thermal challenge, overcome by using a fine-tuned composition of radiation shields and thermal insulators. By coupling the vibration decoupling system only to selected stages of the cryostat with flexible thermal links, reintroduction of cryostat related vibrations into the system is minimized. Further studies in lowering the cross section of said thermal links by increasing the thermal coupling strength, and thus improving their flexibility, have the potential of further improving the achieved vibration levels and/or temperature stability.\\
    Stable operating temperatures below 10\,mK could be achieved in experiment repeatedly after less than two days of cooling time. The remaining vibration levels were shown to be below what is typically achieved in an experimental volume subject to environmental vibrations only. It was demonstrated that the created cryogenic low vibration environment allows indefinite stable operation of TES-based gram-scale calorimeters at peak performance, uninfluenced by pulse tube vibrations. The presented approach can also be applied to detectors with a higher mass. Although an overall larger mass would further improve axial decoupling (see equation \ref{eq:spring_resonance}), for the case of cryogenic calorimeters an increased mass leads to slower pulse time constants \cite{probst1995model}. This can cause a shift of the detector ROI to lower frequencies. Given the measured VDS attenuation above 10\,Hz in axial and 5\,Hz in radial directions, the presented setup is expected to perform comparably in combination with detectors featuring moderately increased pulse time constants. For extremely slow pulses, additional optimization of the thermal links in order to further reduce low-frequency vibrations may be required. As such, the developed vibration decoupling system facilitates the transition to dry dilution refrigerators for the cryogenic detector community without compromising performance due to pulse tube-induced vibrations.
	\paragraph{Funding} 
	The research has been supported by the ERC-StG2018-804228 NUCLEUS, by which the NUCLEUS experiment is funded and the SFB1258 ”Neutrinos and Dark Matter in Astro- and Particle Physics”. NUCLEUS members acknowledge additional funding by the DFG through the  Excellence Cluster ORIGINS, by the P2IO LabEx (ANR-10-LABX-0038) in the framework "Investissements d’Avenir" (ANR-11-IDEX-0003-01) managed by the Agence Nationale de la Recherche (ANR), France and by the Austrian Science Fund (FWF) through the "P 34778-N, ELOISE".
 
    
    \printbibliography 
\end{document}